\documentclass[a4paper,11pt]{article}
\usepackage[T1]{fontenc} % if needed

\makeatletter
\gdef\@fpheader{ }
\gdef\@journal{ }
\makeatother
\RequirePackage{amsmath}
\RequirePackage{amssymb}
\RequirePackage{epsfig}
\RequirePackage{graphicx}
\RequirePackage[numbers,sort&compress]{natbib}
\RequirePackage{color}
\RequirePackage[colorlinks=true
,urlcolor=blue
,anchorcolor=blue
,citecolor=blue
,filecolor=blue
,linkcolor=blue
,menucolor=blue
,pagecolor=blue
,linktocpage=true
,pdfproducer=medialab
,pdfa=true
]{hyperref}

\newif\ifnotoc\notocfalse
\newif\ifemailadd\emailaddfalse
\newif\iftoccontinuous\toccontinuousfalse
\makeatletter
\def\@subheader{\@empty}
\def\@keywords{\@empty}
\def\@abstract{\@empty}
\def\@xtum{\@empty}
\def\@dedicated{\@empty}
\def\@arxivnumber{\@empty}
\def\@collaboration{\@empty}
\def\@collaborationImg{\@empty}
\def\@proceeding{\@empty}
\def\@preprint{\@empty}

\newcommand{\subheader}[1]{\gdef\@subheader{#1}}
\newcommand{\keywords}[1]{\if!\@keywords!\gdef\@keywords{#1}\else%
\PackageWarningNoLine{\jname}{Keywords already defined.\MessageBreak Ignoring last definition.}\fi}
\renewcommand{\abstract}[1]{\gdef\@abstract{#1}}
\newcommand{\dedicated}[1]{\gdef\@dedicated{#1}}
\newcommand{\arxivnumber}[1]{\gdef\@arxivnumber{#1}}
\newcommand{\proceeding}[1]{\gdef\@proceeding{#1}}
\newcommand{\xtumfont}[1]{\textsc{#1}}
\newcommand{\correctionref}[3]{\gdef\@xtum{\xtumfont{#1} \href{#2}{#3}}}
\newcommand\jname{JHEP}
\newcommand\acknowledgments{\section*{Acknowledgments}}

\newcommand\preprint[1]{\gdef\@preprint{\hfill #1}}

\makeatother

\newcommand\note[2][]{%
\if!#1!%
\stepcounter{footnote}\footnotetext{#2}%
\else%
{\renewcommand\thefootnote{#1}%
\footnotetext{#2}}%
\fi}

\makeatletter
\newtoks\auth@toks
\renewcommand{\author}[2][]{%
  \if!#1!%
    \auth@toks=\expandafter{\the\auth@toks#2\ }%
  \else
    \auth@toks=\expandafter{\the\auth@toks#2$^{#1}$\ }%
  \fi
}
\makeatother
\makeatletter
\newtoks\affil@toks\newif\ifaffil\affilfalse
\newcommand{\affiliation}[2][]{%
\affiltrue
  \if!#1!%
    \affil@toks=\expandafter{\the\affil@toks{\item[]#2}}%
  \else
    \affil@toks=\expandafter{\the\affil@toks{\item[$^{#1}$]#2}}%
  \fi
}
\makeatother
\makeatletter
\newtoks\email@toks\newcounter{email@counter}%
\newcommand{\emailAdd}[1]{%
\emailaddtrue%
\ifnum\theemail@counter>0\email@toks=\expandafter{\the\email@toks, \@email{#1}}%
\else\email@toks=\expandafter{\the\email@toks\@email{#1}}%
\fi\stepcounter{email@counter}}
\newcommand{\@email}[1]{\href{mailto:#1}{\tt #1}}
\makeatother

\makeatletter
\newcommand*\collaboration[1]{\gdef\@collaboration{#1}}
\newcommand*\collaborationImg[2][]{\gdef\@collaborationImg{#2}}
\makeatletter
\newcommand\afterLogoSpace{\smallskip}
\newcommand\afterSubheaderSpace{\vskip3pt plus 2pt minus 1pt}
\newcommand\afterProceedingsSpace{\vskip21pt plus0.4fil minus15pt}
\newcommand\afterTitleSpace{\vskip23pt plus0.06fil minus13pt}
\newcommand\afterRuleSpace{\vskip23pt plus0.06fil minus13pt}
\newcommand\afterCollaborationSpace{\vskip3pt plus 2pt minus 1pt}
\newcommand\afterCollaborationImgSpace{\vskip3pt plus 2pt minus 1pt}
\newcommand\afterAuthorSpace{\vskip5pt plus4pt minus4pt}
\newcommand\afterAffiliationSpace{\vskip3pt plus3pt}
\newcommand\afterEmailSpace{\vskip16pt plus9pt minus10pt\filbreak}
\newcommand\afterXtumSpace{\par\bigskip}
\newcommand\afterAbstractSpace{\vskip16pt plus9pt minus13pt}
\newcommand\afterKeywordsSpace{\vskip16pt plus9pt minus13pt}
\newcommand\afterArxivSpace{\vskip3pt plus0.01fil minus10pt}
\newcommand\afterDedicatedSpace{\vskip0pt plus0.01fil}
\newcommand\afterTocSpace{\bigskip\medskip}
\newcommand\afterTocRuleSpace{\bigskip\bigskip}
\newlength{\affiliationsSep}
\newcommand\beforetochook{\pagestyle{myplain}\pagenumbering{roman}}

\DeclareFixedFont\trfont{OT1}{phv}{b}{sc}{11}

\renewcommand\maketitle{
\pagestyle{empty}
\thispagestyle{titlepage}
\setcounter{page}{0}
\noindent{\small\scshape\@fpheader}\@preprint\par

\afterLogoSpace
\if!\@subheader!\else\noindent{\trfont{\@subheader}}\fi
\afterSubheaderSpace
\if!\@proceeding!\else\noindent{\sc\@proceeding}\fi
\afterProceedingsSpace
{\LARGE\flushleft\sffamily\bfseries\@title\par}
\afterTitleSpace
\hrule height 1.5\p@%
\afterRuleSpace
\if!\@collaboration!\else
{\Large\bfseries\sffamily\raggedright\@collaboration}\par
\afterCollaborationSpace
\fi
\if!\@collaborationImg!\else
{\normalsize\bfseries\sffamily\raggedright\@collaborationImg}\par
\afterCollaborationImgSpace
\fi
{\bfseries\raggedright\sffamily\the\auth@toks\par}
\afterAuthorSpace
\ifaffil\begin{list}{}{%
\setlength{\leftmargin}{0.28cm}%
\setlength{\labelsep}{0pt}%
\setlength{\itemsep}{\affiliationsSep}%
\setlength{\topsep}{-\parskip}}
\itshape\small%
\the\affil@toks
\end{list}\fi
\afterAffiliationSpace
\ifemailadd %% if emailadd is true
\noindent\hspace{0.28cm}\begin{minipage}[l]{.9\textwidth}
\begin{flushleft}
\textit{E-mail:} \the\email@toks
\end{flushleft}
\end{minipage}
\else %% if emailaddfalse do nothing
\PackageWarningNoLine{\jname}{E-mails are missing.\MessageBreak Plese use \protect\emailAdd\space macro to provide e-mails.}
\fi
\afterEmailSpace
\if!\@xtum!\else\noindent{\@xtum}\afterXtumSpace\fi
\if!\@abstract!\else\noindent{\renewcommand\baselinestretch{.9}\textsc{Abstract:}}\ \@abstract\afterAbstractSpace\fi
\if!\@keywords!\else\noindent{\textsc{Keywords:}} \@keywords\afterKeywordsSpace\fi
\if!\@arxivnumber!\else\noindent{\textsc{ArXiv ePrint:}} \href{http://arxiv.org/abs/\@arxivnumber}{\@arxivnumber}\afterArxivSpace\fi
\if!\@dedicated!\else\vbox{\small\it\raggedleft\@dedicated}\afterDedicatedSpace\fi
\ifnotoc\else
\iftoccontinuous\else\newpage\fi
\beforetochook\hrule
\tableofcontents
\afterTocSpace
\hrule
\afterTocRuleSpace
\fi
\setcounter{footnote}{0}
\pagestyle{myplain}\pagenumbering{arabic}
} % close the \renewcommand\maketitle{

\renewcommand{\baselinestretch}{1.1}\normalsize
\divide\@tempdima\baselineskip
\@tempcnta=\@tempdima
\skip\@mpfootins = \skip\footins
\renewcommand{\@dotsep}{10000}

\newcommand\ps@myplain{
\pagenumbering{arabic}
\renewcommand\@oddfoot{\hfill-- \thepage\ --\hfill}
\renewcommand\@oddhead{}}
\let\ps@plain=\ps@myplain

\newcommand\ps@titlepage{\renewcommand\@oddfoot{}\renewcommand\@oddhead{}}

\numberwithin{equation}{section}

\renewcommand\section{\@startsection{section}{1}{\z@}%
                                   {-3.5ex \@plus -1.3ex \@minus -.7ex}%
                                   {2.3ex \@plus.4ex \@minus .4ex}%
                                   {\normalfont\large\bfseries}}
\renewcommand\subsection{\@startsection{subsection}{2}{\z@}%
                                   {-2.3ex\@plus -1ex \@minus -.5ex}%
                                   {1.2ex \@plus .3ex \@minus .3ex}%
                                   {\normalfont\normalsize\bfseries}}
\renewcommand\subsubsection{\@startsection{subsubsection}{3}{\z@}%
                                   {-2.3ex\@plus -1ex \@minus -.5ex}%
                                   {1ex \@plus .2ex \@minus .2ex}%
                                   {\normalfont\normalsize\bfseries}}
\renewcommand\paragraph{\@startsection{paragraph}{4}{\z@}%
                                   {1.75ex \@plus1ex \@minus.2ex}%
                                   {-1em}%
                                   {\normalfont\normalsize\bfseries}}
\renewcommand\subparagraph{\@startsection{subparagraph}{5}{\parindent}%
                                   {1.75ex \@plus1ex \@minus .2ex}%
                                   {-1em}%
                                   {\normalfont\normalsize\bfseries}}

\def\fnum@figure{\textbf{\figurename\nobreakspace\thefigure}}
\def\fnum@table{\textbf{\tablename\nobreakspace\thetable}}

\long\def\@makecaption#1#2{%
  \vskip\abovecaptionskip
  \sbox\@tempboxa{\small #1. #2}%
  \ifdim \wd\@tempboxa >\hsize
    \small #1. #2\par
  \else
    \global \@minipagefalse
    \hb@xt@\hsize{\hfil\box\@tempboxa\hfil}%
  \fi
  \vskip\belowcaptionskip}

\renewenvironment{thebibliography}[1]{%
\begin{oldthebibliography}{#1}%
\small%
\raggedright%
\setlength{\itemsep}{5pt plus 0.2ex minus 0.05ex}%
}%
{%
\end{oldthebibliography}%
}

\begin{document}

%%%%%%%%%%%%%%%%%%±êÌâÒ³%%%%%%%%%%%%%%%%%%%%%%%%%%%%%&&&&&&&&&&&&&&&&&&&&&&

%\title{\boldmath Renormalization for singular-potential scattering ${}^{\dag}$ \footnotetext{$\dag$ This is an
%enlarged version of the corresponding content in Ref. \cite{li2016scattering}.}}%

\renewcommand{\thefootnote}{\fnsymbol{footnote}}%°Ñ½Å×¢±àºÅ±ä³É·ÇÊý×ÖÇé¿ö

\title{\boldmath Renormalization for singular-potential scattering\footnote{ This is an
enlarged version of the corresponding content in Ref. \cite{li2016scattering}.}}%

% more complex case: 4 authors, 3 institutions, 2 footnotes
\author[a]{Wen-Du Li}
\author[a,b,*]{and Wu-Sheng Dai}\note{daiwusheng@tju.edu.cn.}

% The "\note" macro will give a warning: "Ignoring empty anchor..."
% you can safely ignore it.

\affiliation[a]{Department of Physics, Tianjin University, Tianjin 300072, P.R. China}
\affiliation[b]{LiuHui Center for Applied Mathematics, Nankai University \& Tianjin University, Tianjin 300072, P.R. China}
%\affiliation[c]{DP School}

% e-mail addresses: one for each author, in the same order as the authors
%\emailAdd{Ccc@one.edu.cn}
%\emailAdd{second@asas.edu}
%\emailAdd{daiwusheng@tju.edu.cn}
%\emailAdd{fourth@one.univ}

%\title{\boldmath A title with some math: $x=1$}
%% %simple case: 2 authors, same institution
%% \author{A. Uthor}
%% \author{and A. Nother Author}
%% \affiliation{Institution,\\Address, Country}

% more complex case: 4 authors, 3 institutions, 2 footnotes
%\author[a,b,1]{F. Irst,\note{Corresponding author.}}
%\author[c]{S. Econd,}
%\author[a,2]{T. Hird\note{Also at Some University.}}
%\author[a,2]{and Fourth}

% The "\note" macro will give a warning: "Ignoring empty anchor..."
% you can safely ignore it.

%\affiliation[a]{One University,\\some-street, Country}
%\affiliation[b]{Another University,\\different-address, Country}
%\affiliation[c]{A School for Advanced Studies,\\some-location, Country}

% e-mail addresses: one for each author, in the same order as the authors

%\emailAdd{first@one.univ}
%\emailAdd{second@asas.edu}
%\emailAdd{third@one.univ}
%\emailAdd{fourth@one.univ}

%\date{date}

\abstract{In the calculation of quantum-mechanical singular-potential scattering, one
encounters divergence. We suggest three renormalization schemes, dimensional
renormalization,\ analytic continuation approach, and minimal-subtraction
scheme to remove the divergence.
}
%\keywords{***}

\maketitle
\flushbottom
%%%%%%%%%%%%%%%%%%±êÌâÒ³½áÊø%%%%%%%%%%%%%%%%%%%%%%%%%%%%%&&&&&&&&&&&&&&&&&&&

%%%%%%%%%%ÕýÎÄ¿ªÊ¼

\section{Introduction}

Once a potential is a singular potential, one encounters divergences. In order
to remove the divergence, we need a renormalization procedure. In this paper,
we suggest three renormalization schemes, dimensional
renormalization,\ analytic continuation approach, and minimal-subtraction
scheme for removing the divergence in singular-potential scattering.

In order to illustrate the validity of dimensional renormalization, we
consider the Born approximation.

To demonstrate the validity of renormalization, by taking the Lennard-Jones
potential as an example, through three different approaches, we perform
renormalization treatments. It will be shown that the results obtained by
these three different approaches are the same.

In the following, we take the Lennard-Jones potential $V\left(  r\right)
=\eta\frac{6}{m-6}\left(  \alpha\frac{1}{r^{m}}-\frac{\beta}{6}\frac{m}{r^{6}%
}\right)  $ as an example to testify the validity of the dimensional
renormalization through a comparison with other two renormalization treatments.

The Lennard-Jones potential with $m=12$ reads%
\begin{equation}
V\left(  r\right)  =\eta\left(  \frac{\alpha}{r^{12}}-\frac{2\beta}{r^{6}%
}\right)  .\label{LJ}%
\end{equation}
A divergence appears in the first-order Born approximation of $s$-wave
scattering phase shift \cite{friedrich2013scattering}:
\begin{equation}
\delta_{0}\left(  k\right)  =-\frac{\pi}{2}\int_{0}^{\infty}rdr\eta\left(
\frac{\alpha}{r^{12}}-\frac{2\beta}{r^{6}}\right)  J_{1/2}^{2}\left(
kr\right)  ,\label{LJs0}%
\end{equation}
where $J_{\nu}\left(  z\right)  $ is the Bessel function of the first kind. In
the following, we remove this divergence through three renormalization
treatments, dimensional renormalization, the analytic continuation approach,
and the minimal-subtraction scheme, to show the validity.

The dimensional renormalization approach is introduced in section \ref{DR}.
The analytic continuation approach is introduced in section \ref{AC}. A
minimal-subtraction scheme is introduced in section \ref{MS}. The conclusion
is given in section \ref{Conclusions}.

\subsection{Dimensional renormalization \label{DR}}

Using the $n$-dimensional Born approximation \cite{graham2009spectral}, we
have the first-order scattering phase shift,%
\begin{equation}
\delta_{l}^{\left(  n\right)  }\left(  k\right)  =-\frac{\pi}{2}\int%
_{0}^{\infty}rdrV\left(  r\right)  J_{n/2+l-1}^{2}\left(  kr\right)  .
\label{ndborn}%
\end{equation}
The $n$-dimensional $s$-wave phase shift of the Lennard-Jones potential
(\ref{LJ}) can be obtained by performing the integral directly,%
\begin{align}
\delta_{0}^{\left(  n\right)  }\left(  k\right)   &  =-\frac{\pi}{2}\int%
_{0}^{\infty}rdr\eta\left(  \frac{\alpha}{r^{12}}-\frac{2\beta}{r^{6}}\right)
J_{n/2-1}^{2}\left(  kr\right) \nonumber\\
&  =-\frac{63\pi\alpha\eta k^{10}\Gamma\left(  n/2-6\right)  }{2^{10}%
\Gamma\left(  n/2+5\right)  }+\frac{3\pi\beta\eta k^{4}\Gamma\left(
n/2-3\right)  }{2^{4}\Gamma\left(  n/2+2\right)  }. \label{delta-d}%
\end{align}
Then directly putting $n=3$ gives%
\begin{equation}
\delta_{0}^{\left(  3\right)  }\left(  k\right)  =\frac{2}{155\,925}\pi
\alpha\eta k^{10}+\frac{2}{15}\pi\beta\eta k^{4}.
\end{equation}
This is a finite result.

\subsection{Analytic continuation approach \label{AC}}

As comparison, we now use another renormalization treatment to remove the
divergence, which is based on analytic continuation.

The integral $I=\int_{0}^{\infty}f\left(  x\right)  dx$ will diverge, if the
expansion of $f\left(  x\right)  $ at $x=0$ has negative-power terms. In order
to remove the divergence, we use the analytic continuation technique.
Concretely, we rewrite the integral as \cite{zeidler2008quantum}%
\begin{align}
I  &  =\int_{0}^{\infty}f\left(  x\right)  dx\nonumber\\
&  =\int_{0}^{1}\left(  f\left(  x\right)  -\sum_{n=2}^{N}a_{n}\frac{1}{x^{n}%
}\right)  dx+\sum_{n=2}^{N}a_{n}\frac{x^{1-n}}{1-n}+\int_{1}^{\infty}f\left(
x\right)  dx, \label{I-Zeidler}%
\end{align}
where\ $a_{n}$ is the expansion coefficient and $N$ equals the highest
negative power of the expansion of $f\left(  x\right)  $. Here the integral is
split into to\ two parts: $\int_{1}^{\infty}dx$ and $\int_{0}^{1}dx$. The
integral $\int_{1}^{\infty}dx$ is well defined. The divergence encountered in
the integral $\int_{0}^{1}dx$ is removed by $\sum_{n=2}^{N}a_{n}\frac{1}%
{x^{n}}$. The basis of eq. (\ref{I-Zeidler}) is essentially analytic continuation.

First, split the integral in Eq. (\ref{LJs0}) into two parts:%

\begin{equation}
\delta_{0}\left(  k\right)  =\left[  \delta_{0}\left(  k\right)  \right]
_{0}^{\epsilon}+\left[  \delta_{0}\left(  k\right)  \right]  _{\epsilon
}^{\infty},
\end{equation}
where%
\begin{align}
\left[  \delta_{0}\left(  k\right)  \right]  _{0}^{\epsilon}  &  =-\frac{\pi
}{2}\int_{0}^{\epsilon}rdr\eta\left(  \frac{\alpha}{r^{12}}-\frac{2\beta
}{r^{6}}\right)  J_{1/2}^{2}\left(  kr\right)  ,\label{0-ep}\\
\left[  \delta_{0}\left(  k\right)  \right]  _{\epsilon}^{\infty}  &
=-\frac{\pi}{2}\int_{\epsilon}^{\infty}rdr\eta\left(  \frac{\alpha}{r^{12}%
}-\frac{2\beta}{r^{6}}\right)  J_{1/2}^{2}\left(  kr\right)  , \label{inf-ep}%
\end{align}
where $\epsilon$ is a finite number.

The integral in Eq. (\ref{inf-ep}) can be performed directly,%
\begin{align}
\left[  \delta_{0}\left(  k\right)  \right]  _{\epsilon}^{\infty}  &
=-\frac{\alpha\eta}{22k\epsilon^{11}}+\frac{\beta\eta}{5k\epsilon^{5}%
}-\operatorname*{Si}(2k\epsilon)\left(  \frac{4\alpha\eta k^{10}}%
{155925}+\frac{4\beta\eta k^{4}}{15}\right)  +\frac{2\pi\alpha\eta k^{10}%
}{155925}+\frac{2\pi\beta\eta k^{4}}{15}\nonumber\\
&  +\sin\left(  2k\epsilon\right)  \left[  -\frac{\alpha\eta}{110\epsilon
^{10}}+\frac{\alpha\eta k^{2}}{1980\epsilon^{8}}-\frac{\alpha\eta k^{4}%
}{20790\epsilon^{6}}+\left(  \frac{\alpha\eta k^{6}}{103950}+\frac{\beta\eta
}{10}\right)  \frac{1}{\epsilon^{4}}-\left(  \frac{\alpha\eta k^{8}}%
{155925}-\frac{\beta\eta k^{2}}{15}\right)  \frac{1}{\epsilon^{2}}\right]
\nonumber\\
&  +\cos\left(  2k\epsilon\right)  \left[  \frac{\alpha\eta}{22k\epsilon^{11}%
}-\frac{\alpha\eta k}{495\epsilon^{9}}+\frac{\alpha\eta k^{3}}{6930\epsilon
^{7}}-\left(  \frac{\alpha\eta k^{5}}{51975}+\frac{\beta\eta}{5k}\right)
\frac{1}{\epsilon^{5}}\right. \nonumber\\
&  +\left.  \left(  \frac{\alpha\eta k^{7}}{155925}+\frac{\beta\eta k}%
{15}\right)  \frac{1}{\epsilon^{3}}-\left(  \frac{2\alpha\eta k^{9}}%
{155925}-\frac{2\beta\eta k^{3}}{15}\right)  \frac{1}{\epsilon}\right]  ,
\end{align}
where $\operatorname*{Si}(z)$ is the sine integral \cite{olver2010nist}.

Now we deal with $\left[  \delta_{0}\left(  k\right)  \right]  _{0}^{\epsilon
}$.

Expanding the integrand in $\left[  \delta_{0}\left(  k\right)  \right]
_{0}^{\epsilon}$ around $r=0$ gives%
\begin{equation}
r\eta\left(  \frac{\alpha}{r^{12}}-\frac{2\beta}{r^{6}}\right)  J_{1/2}%
^{2}\left(  kr\right)  =D\left(  r\right)  ,
\end{equation}
where%
\begin{equation}
D\left(  r\right)  =\frac{2\alpha\eta k}{\pi r^{10}}-\frac{2\alpha\eta k^{3}%
}{3\pi r^{8}}+\frac{4\alpha\eta k^{5}}{45\pi r^{6}}-\left(  \frac{2\alpha\eta
k^{7}}{315\pi}+\frac{4\beta\eta k}{\pi}\right)  \frac{1}{r^{4}}+\left(
\frac{4\alpha\eta k^{9}}{14175\pi}+\frac{4\beta\eta k^{3}}{3\pi}\right)
\frac{1}{r^{2}}+\cdots.
\end{equation}
According to Eq. (\ref{I-Zeidler}), subtracting $D\left(  r\right)  $ from the
integral in Eq. (\ref{0-ep}) gives%
\begin{align}
&  -\frac{\pi}{2}\int_{0}^{\epsilon}dr\left[  r\eta\left(  \frac{\alpha
}{r^{12}}-\frac{2\beta}{r^{6}}\right)  J_{1/2}^{2}\left(  kr\right)  -D\left(
r\right)  \right] \nonumber\\
&  =\frac{\alpha\eta}{22k\epsilon^{11}}-\frac{\alpha\eta k}{9\epsilon^{9}%
}+\frac{\alpha\eta k^{3}}{21\epsilon^{7}}-\left(  \frac{\beta\eta}{5k}%
+\frac{2\alpha\eta k^{5}}{225}\right)  \frac{1}{\epsilon^{5}}+\left(
\frac{\alpha\eta k^{7}}{945}+\frac{2\beta\eta k}{3}\right)  \frac{1}%
{\epsilon^{3}}-\left(  \frac{2\alpha\eta k^{9}}{14175}+\frac{2\beta\eta k^{3}%
}{3}\right)  \frac{1}{\epsilon}\nonumber\\
&  +4\operatorname*{Si}\left(  2k\epsilon\right)  \left(  \frac{\alpha\eta
k^{10}}{155925}+\frac{\beta\eta k^{4}}{15}\right) \nonumber\\
&  +\sin\left(  2k\epsilon\right)  \left[  \frac{\alpha\eta}{110\epsilon^{10}%
}-\frac{\alpha\eta k^{2}}{1980\epsilon^{8}}+\frac{\alpha\eta k^{4}%
}{20790\epsilon^{6}}-\left(  \frac{\alpha\eta k^{6}}{103950}+\frac{\beta\eta
}{10}\right)  \frac{1}{\epsilon^{4}}+\left(  \frac{\alpha\eta k^{8}}%
{155925}+\frac{\beta\eta k^{2}}{15}\right)  \frac{1}{\epsilon^{2}}\right]
\nonumber\\
&  +\cos\left(  2k\epsilon\right)  \left[  -\frac{\alpha\eta}{22k\epsilon
^{11}}+\frac{\alpha\eta k}{495\epsilon^{9}}-\frac{\alpha\eta k^{3}%
}{6930\epsilon^{7}}+\left(  \frac{\alpha\eta k^{5}}{51975}+\frac{\beta\eta
}{5k}\right)  \frac{1}{\epsilon^{5}}\right. \nonumber\\
&  -\left.  \left(  \frac{\alpha\eta k^{7}}{155925}+\frac{\beta\eta k}%
{15}\right)  \frac{1}{\epsilon^{3}}+\left(  \frac{2\alpha\eta k^{9}}%
{155925}+\frac{2\beta\eta k^{3}}{15}\right)  \frac{1}{\epsilon}\right]  .
\end{align}
Introducing
\begin{equation}
D_{s}\left(  r\right)  =\frac{2\alpha\eta k}{\pi r^{10s}}-\frac{2\alpha\eta
k^{3}}{3\pi r^{8s}}+\frac{4\alpha\eta k^{5}}{45\pi r^{6s}}-\left(
\frac{2\alpha\eta k^{7}}{315\pi}+\frac{4\beta\eta k}{\pi}\right)  \frac
{1}{r^{4s}}+\left(  \frac{4\alpha\eta k^{9}}{14175\pi}+\frac{4\beta\eta k^{3}%
}{3\pi}\right)  \frac{1}{r^{2s}}+\cdots
\end{equation}
with $\left.  D_{s}\left(  r\right)  \right\vert _{s=1}=D\left(  r\right)  $.

Integrating $D_{s}\left(  r\right)  $ and\ then putting $s=1$ give
\begin{align}
&  \left.  -\frac{\pi}{2}\int_{0}^{\epsilon}drD_{s}\left(  r\right)
\right\vert _{s=1}\nonumber\\
&  =\left.  \frac{\eta k}{14175}\left[  \frac{14175\alpha}{\left(
10s-1\right)  \epsilon^{10s-1}}-\frac{4725\alpha k^{2}}{\left(  8s-1\right)
\epsilon^{8s-1}}+\frac{630\alpha k^{4}}{\left(  6s-1\right)  \epsilon^{6s-1}%
}-\frac{45\left(  630\beta+\alpha k^{6}\right)  }{\left(  4s-1\right)
\epsilon^{4s-1}}+\frac{2k^{2}\left(  4725\beta+\alpha k^{6}\right)  }{\left(
2s-1\right)  \epsilon^{2s-1}}\right]  \right\vert _{s=1}\nonumber\\
&  =\frac{\alpha\eta k}{9\epsilon^{9}}-\frac{\alpha\eta k^{3}}{21\epsilon^{7}%
}+\frac{2\alpha\eta k^{5}}{225\epsilon^{5}}-\left(  \frac{\alpha\eta k^{7}%
}{945}+\frac{2\beta\eta k}{3}\right)  \frac{1}{\epsilon^{3}}+\left(
\frac{2\beta\eta k^{3}}{3}+\frac{2\alpha\eta k^{9}}{14175}\right)  \frac
{1}{\epsilon}.
\end{align}
Then we arrive at%
\begin{align}
\delta_{0}\left(  k\right)   &  =-\frac{\pi}{2}\int_{0}^{\epsilon}dr\left[
r\eta\left(  \frac{\alpha}{r^{12}}-\frac{2\beta}{r^{6}}\right)  J_{1/2}%
^{2}\left(  kr\right)  -D\left(  r\right)  \right]  +\left(  \left.
-\frac{\pi}{2}\int_{0}^{\epsilon}drD_{s}\left(  r\right)  \right\vert
_{s=1}\right) \nonumber\\
&  -\frac{\pi}{2}\int_{\epsilon}^{\infty}rdr\eta\left(  \frac{\alpha}{r^{12}%
}-\frac{2\beta}{r^{6}}\right)  J_{1/2}^{2}\left(  kr\right) \nonumber\\
&  =\frac{2}{155\,925}\pi\alpha\eta k^{10}+\frac{2}{15}\pi\beta\eta k^{4}.
\end{align}
This result agrees with that given by dimensional renormalization.

\subsection{Minimal-subtraction scheme \label{MS}}

The minimal-subtraction scheme is simply to remove the poles in divergent
quantities \cite{peskin1995introduction}.

First directly cut off the lower limit of the integral in Eq. (\ref{LJs0}):%
\begin{equation}
\delta_{0}\left(  k,\epsilon\right)  =-\frac{\pi}{2}\int_{\epsilon}^{\infty
}rdr\eta\left(  \frac{\alpha}{r^{12}}-\frac{2\beta}{r^{6}}\right)  J_{1/2}%
^{2}\left(  kr\right)  . \label{deltae}%
\end{equation}
For $\epsilon>0$, the integral is convergent.

Performing the integral in Eq. (\ref{deltae}) gives
\begin{align}
\delta_{0}\left(  k,\epsilon\right)   &  =-\frac{\alpha\eta}{22k\epsilon^{11}%
}+\frac{\beta\eta}{5k\epsilon^{5}}+\frac{2\pi\alpha\eta k^{10}}{155925}%
+\frac{2\pi\beta\eta k^{4}}{15}-\operatorname*{Si}(2k\epsilon)\left(
\frac{4\alpha\eta k^{10}}{155925}+\frac{4\beta\eta k^{4}}{15}\right)
\nonumber\\
&  +\cos\left(  2k\epsilon\right)  \left[  \frac{\alpha\eta}{22k\epsilon^{11}%
}-\frac{\alpha\eta k}{495\epsilon^{9}}+\frac{\alpha\eta k^{3}}{6930\epsilon
^{7}}-\frac{\alpha\eta k^{5}}{51975\epsilon^{5}}-\frac{\beta\eta}%
{5k\epsilon^{5}}\right. \nonumber\\
&  \left.  +\left(  \frac{\alpha\eta k^{7}}{155925}+\frac{\beta\eta k}%
{15}\right)  \frac{1}{\epsilon^{3}}-\left(  \frac{2\alpha\eta k^{9}}%
{155925}+\frac{2\beta\eta k^{3}}{15}\right)  \frac{1}{\epsilon}\right]
\nonumber\\
&  +\sin\left(  2k\epsilon\right)  \left[  -\frac{\alpha\eta}{110\epsilon
^{10}}+\frac{\alpha\eta k^{2}}{1980\epsilon^{8}}-\frac{\alpha\eta k^{4}%
}{20790\epsilon^{6}}+\left(  \frac{\alpha\eta k^{6}}{103950}+\frac{\beta\eta
}{10}\right)  \frac{1}{\epsilon^{4}}-\left(  \frac{\alpha\eta k^{8}}%
{155925}+\frac{\beta\eta k^{2}}{15}\right)  \frac{1}{\epsilon^{2}}\right]  .
\end{align}
Expanding $\delta_{0}\left(  k,\epsilon\right)  $ around $\epsilon=0$ gives%
\begin{equation}
\delta_{0}\left(  k,\epsilon\right)  =-\frac{\alpha\eta k}{9\epsilon^{9}%
}+\frac{\alpha\eta k^{3}}{21\epsilon^{7}}-\frac{2\alpha\eta k^{5}}%
{225\epsilon^{5}}+\frac{\alpha\eta k^{7}+630\beta\eta k}{945\epsilon^{3}%
}-\frac{2\alpha\eta k^{9}+9450\beta\eta k^{3}}{14175\epsilon}+\frac{2\pi
\alpha\eta k^{10}+20\,790\pi\beta\eta k^{4}}{155925}+O\left(  \epsilon\right)
.
\end{equation}
Take $\epsilon\rightarrow0$ and, according to the minimal-subtraction scheme,
dropping out the terms that diverge when $\epsilon\rightarrow0$ give%
\begin{equation}
\delta_{0}\left(  k\right)  =\frac{2}{155\,925}\pi\alpha\eta k^{10}+\frac
{2}{15}\pi\beta\eta k^{4}.
\end{equation}

\section{Conclusions \label{Conclusions}}

The above three renormalized results agree with each other demonstrates the
validity of these three renormalization schemes.

In dimensional renormalization scheme, we need an arbitrary dimensional theory
in which the value of spatial dimension $n$ appears as a renormalization
parameter, like that in dimensional renormalization in quamtum field theory
\cite{graham2009spectral,pang2012relation,li2015heat,rahi2009scattering,forrow2012variable,bimonte2012exact,dai2009number,dai2010approach}%
. Moreover, the above schemes can also be examined through exact solutions,
e.g. \cite{Li2016exact}. The renormalization treatment introduced in the
present paper should be applied to more general scattering problems
\cite{liu2014scattering}.

%\appendix
%\section{Some title}
%Please always give a title also for appendices.

\acknowledgments

We are very indebted to Dr G. Zeitrauman for his encouragement. This work is supported in part by NSF of China under Grant
No. 11575125 and No. 11375128.

%\acknowledgments%ÖÂл
%%%%%%%%%%ÕýÎĽáÊø

%\begin{thebibliography}{99}

%\end{thebibliography}\endgroup

%\bibitem{a}
%Author, \emph{Title}, \emph{J. Abbrev.} {\bf vol} (year) pg.

%\bibitem{b}
%Author, \emph{Title},
%arxiv:1234.5678.

%\bibitem{c}
%Author, \emph{Title},
%Publisher (year).

% Please avoid comments such as "For a review'', "For some examples",
% "and references therein" or move them in the text. In general,
% please leave only references in the bibliography and move all
% accessory text in footnotes.

% Also, please have only one work for each \bibitem.

%\end{thebibliography}

\providecommand{\href}[2]{#2}\begingroup\raggedright\endgroup

%\bibliographystyle{JHEP}
%\bibliography{refs}% Produces the bibliography via BibTeX.

\end{document}